\documentclass[pra,aps,twocolumn,superscriptaddress,showpacs,amsmath,amssymb,floatfix]{revtex4}

\usepackage{comment}
\usepackage{graphicx}
\usepackage{srcltx}
\usepackage{amssymb}

\begin{document}
\title{Subdiffusion of nonlinear waves in quasiperiodic potentials}
\author{M. Larcher}
\affiliation{INO-CNR BEC Center and Dipartimento di Fisica, Universit\`a di Trento, 38123 Povo, Italy}
\author{T. V. Laptyeva}
\affiliation{Max Planck Institute for the Physics of Complex Systems, N\"othnitzer Str. 38, D-01187 Dresden, Germany}
\author{J. D. Bodyfelt}
\affiliation{Max Planck Institute for the Physics of Complex Systems, N\"othnitzer Str. 38, D-01187 Dresden, Germany}
\author{F. Dalfovo}
\affiliation{INO-CNR BEC Center and Dipartimento di Fisica, Universit\`a di Trento, 38123 Povo, Italy}
\author{M. Modugno}
\affiliation{IKERBASQUE, Basque Foundation for Science, 48011 Bilbao, Spain}
\affiliation{Department of Theoretical Physics and History of Science, Universidad del Pais Vasco UPV/EHU, 48080 Bilbao, Spain}
\author{S. Flach}
\affiliation{Max Planck Institute for the Physics of Complex Systems, N\"othnitzer Str. 38, D-01187 Dresden, Germany}
\affiliation{New Zealand Institute for Advanced Study, Massey University, Auckland, New Zealand}

\begin{abstract}
We study the time evolution of wave packets in one-dimensional quasiperiodic lattices which localize linear waves.  Nonlinearity (related to two-body interactions) has a destructive effect on localization, as recently observed for interacting atomic condensates [Phys. Rev. Lett. {\bf 106}, 230403 (2011)]. We extend the analysis of the characteristics of the subdiffusive dynamics to large temporal and spatial scales. Our results for the second moment $m_2$ consistently reveal an asymptotic $m_2 \sim t^{1/3}$ and intermediate $m_2 \sim t^{1/2}$ laws. At variance to purely random systems [Europhys. Lett. {\bf 91}, 30001 (2010)], the fractal gap structure of the linear wave spectrum strongly favors intermediate self-trapping events. Our findings give a new dimension to the theory of wave packet spreading in localizing environments.
\end{abstract}
\maketitle

\section{Introduction}
In one dimension, a wave packet of noninteracting particles subject to a random potential does not diffuse because of Anderson localization, due to the exponential localization of the  eigenstates of the underlying Hamiltonian \cite{anderson1958, kramer1993}. Instead the presence of interactions is expected to act against localization, though the actual mechanism may be highly nontrivial and may depend on the type of disorder and interaction. This is a problem of fundamental importance for many systems in different contexts. From the theoretical side, it has been largely studied by using discrete lattices. The interaction is often included by means of mean-field theories, where it enters in the form of a nonlinear local term in a nonlinear Schr\"odinger-like equation 
\cite{kopidakis2008,pikovsky2008,garcia2009,flach2009,skokos2009,
veksler2009,flach2010,skokos2010,bodyfelt2010,laptyeva2010,bodyfelt2011}.

Numerical simulations of wave packets propagating in a random potential -- with the interaction included via a nonlinear mean-field term -- showed that the presence of interaction indeed destroys localization 
and leads to a subdiffusive growth of the second moment of the wave  packet in time as $t^\gamma$ \cite{pikovsky2008,garcia2009,flach2009,skokos2009,veksler2009, skokos2010,laptyeva2010,bodyfelt2011}. In particular it was predicted that at large $t$, the coefficient $\gamma$ should converge to $1/3$ in a regime of so-called ``weak chaos'', as opposed to normal diffusion where $\gamma=1$ and the wave packet width grows as $\sqrt{t}$. A transient regime of ``strong chaos'' was also identified, where $\gamma=1/2$ \cite{flach2010,laptyeva2010,bodyfelt2011}. The occurrence of these different regimes can be predicted by comparing the nonlinear frequency shift introduced by the expanding wave packet to the typical energy scales of linear spectra of random models. 

Exponential localization for noninteracting quantum particles (or linear waves) can be found also in systems which are not truly disordered. An example is provided by quasiperiodic potentials, which are of great interest by themselves \cite{hiramoto1992,aubry1980}. These systems can be considered to lie between the two extreme cases of a perfectly periodic system and a pure random potential. By tuning the parameters of a 
quasiperiodic system, the localization properties can change dramatically -- from having all extended states to all localized states. In recent years, exponential localization has been observed with light propagating in quasiperiodic photonic lattices \cite{lahini2009}, as well as with ultracold atoms propagating in a bichromatic optical lattice \cite{roati2008}. Notably in both cases the inclusion of interaction is experimentally feasible, by using a Kerr medium for light and tuning the scattering length by means of a suitable magnetic field for atoms.

Numerical simulations studying nonlinear dynamics of wave packets have been performed also in the case of quasiperiodic systems \cite{johansson1995,ng2007,larcher2009,zhang2011}. In particular for exponentially localized linear waves, nonlinearity yields subdiffusive spreading of wave packets as well \cite{larcher2009}. However, there are clear indications that the coefficient $\gamma$, at least at finite spreading times, is significantly larger than the one observed in random systems. Nonlinear effects have been also studied in experiments using ultracold atoms and light propagating in photonic lattices. In both cases it has been shown that nonlinearity acts against localization \cite{lahini2009, lucioni2011}.

The purpose of the present work is to clarify the details of the spreading mechanism leading to the destruction of the localization in quasiperiodic systems, and to address differences and similarities 
between quasiperiodic and purely random potentials. We extend and refine previous numerical investigations by pushing the simulations to much longer times, thus allowing for the identification of the strong and weak chaos regimes in quasiperiodic systems and compare the situation with known properties of purely random systems. For this purpose, we use two different models; namely, a discrete nonlinear Schr\"odinger equation (DNLS) and a quasiperiodic version of the quartic Klein-Gordon (KG) lattice model.

A key result of the present work is that a regime of weak chaos is indeed observed in the long time spreading of nonlinear wave packets propagating in quasiperiodic systems; in particular we find that the asymptotic value of the spreading coefficient $\gamma$ is $1/3$ as in purely random systems, thus showing that this behaviour is rather general and model independent. Another similarity with purely random systems is the occurrence of self-trapping: when the nonlinear interaction is large enough to shift the mode frequencies so strongly that they are tuned out of resonance with all nonexcited neighbouring modes, a part of the wave packet remains spatially localized \cite{kopidakis2008,flach2009,larcher2009}. However as opposed to the random system, in the quasiperiodic case partial self-trapping is also possible for weaker nonlinearities. This is due to the complexity of the linear wave spectrum which exhibits a fractal gap structure of sub-bands. Self-trapping gives rise to transient spreading regimes characterized by an intermediate large exponent $\gamma$; we call this effect ``overshooting''.  Finally, we have also observed signatures of strong chaos, but detection of this regime is difficult in quasiperiodic systems, since it is often masked by overshooting and partial self-trapping, which occur on the same temporal scales.  

\section{Models}
We consider two different models: the first is the one-dimensional (1D) quasiperiodic discrete nonlinear Schr\"odinger equation (DNLS), defined by the Hamiltonian
\begin{equation}
H_{\rm DNLS}=\sum_j \left[-\psi_{j+1}\psi_j^*-\psi_{j-1}^*\psi_j +V_j|\psi_j|^2 + \frac{\beta}{2}|\psi_j|^4 \right]\, ,\label{eq:H_DNLS}
\end{equation}
where $j$ labels the lattice sites and $V_j=\lambda\cos(2\pi\alpha j+\varphi)$. The quantity $|\psi_j|^2$ gives the probability to find a particle at site $j$. The first term in Eq.~(\ref{eq:H_DNLS}) describes the hopping between nearest neighbouring sites, the second term describes the quasiperiodic on-site energy, while the last term represents the mean-field interaction energy and introduces the nonlinearity. The 
key parameters of this Hamiltonian are the strength of the quasiperiodic potential $\lambda$ (for $\lambda=0$ the lattice is periodic with period $1$), the strength of the nonlinearity $\beta$ (for $\beta=0$ the 
particles are non-interacting), and the irrational number $\alpha$ which causes the underlying potential to be quasiperiodic.  In fact, when $\alpha$ is irrational the cosine adds a second periodicity which is incommensurate with respect to the underlying periodicity given by the discreteness of the system. Let us note that, without any loss of generality, one can always choose $-0.5<\alpha \le0.5$ since Eq.~(\ref{eq:H_DNLS}) and the subsequent Eq.~(\ref{eq:H_KG}) are invariant under a shift of $\alpha$ by an integer number. Choosing the value of $\alpha$ in this interval has the additional advantage that the wavenumber associated to $V_j$, $k_\alpha=2 \pi \alpha$, rests in the Brillouin zone of the underlying discrete lattice. As a convenient choice for this work, we use $\alpha=(\sqrt{5}-3)/{2}$ \cite{alpha}.
The phase $\varphi$ is a phase shift between the two lattices. The equations of motion associated with Eq.~(\ref{eq:H_DNLS}) are  $i{\partial \psi_j}/{\partial t}={\partial H}/{\partial \psi_j^*}$, or 
\begin{eqnarray}
i\frac{\partial \psi_j}{\partial t} = -(\psi_{j+1}+\psi_{j-1})+V_j\psi_{j}+\beta|\psi_j|^2\psi_j \, . \label{eq:DNLS}
\end{eqnarray}
The above set of equations conserve the energy of Eq.~(\ref{eq:H_DNLS}), as well as the total norm $S=\sum_j \left|\psi_j\right|^2$ of the initial wave packet (we always assume $S=1$).
The set can be used for a wide range of applications, including ultracold atoms expanding in optical lattices \cite{roati2008,modugno2009,larcher2009,trombettoni2001} and light propagating in photonic lattices \cite{lahini2009,christodolides2003}. 

The second model is a quasiperiodic version of the quartic Klein-Gordon (KG) lattice, given by
\begin{equation}
H_{\rm KG}=\frac{1}{2}\sum_j \left[ p_j^2+\tilde{V_j}u_j^2+\frac{1}{2}u_j^4+\frac{1}{2\lambda}(u_{j+1}-u_j)^2\right] ,\label{eq:H_KG}
\end{equation}
where $u_j$ and $p_j$ are the generalized coordinates and momenta on the site $j$ and $\tilde{V_j}=1+(1/2)\cos(2\pi\alpha j+\varphi)$. The energy associated with lattice site $j$ is 
\begin{equation}
\mathcal{E}_j = \frac{p_j^2}{2}+\frac{\tilde{V_j}u_j^2}{2}+\frac{u_j^4}{4}+\frac{(u_{j+1}-u_j)^2}{8 \lambda} +\frac{(u_{j-1}-u_j)^2}{8 \lambda} .\label{eq:on-site_energy}
\end{equation}
The equations of motion are generated by $\partial^2 u_j/\partial t^2=-\partial H/\partial u_j$, yielding
\begin{equation}
\frac{\partial^2 u_j}{\partial t^2}=-\tilde{V_j}u_j-u_j^3+\frac{1}{2\lambda}(u_{j+1}+u_{j-1}-2 u_j) \, .\label{eq:KG}
\end{equation}
This set of equations conserve the total energy $\mathcal{H}=\sum_j \mathcal{E}_j \geq 0$ only. The KG model has also been extensively studied, since it can give a simple description of the nondissipative dynamics of anharmonic optical lattice vibrations in molecular crystals \cite{ovchinnikov2001}.

The total energy of the system $\mathcal{H}$ serves as a control parameter of nonlinearity, analogous to $\beta$ for the DNLS model. In fact, for small amplitudes the equation of the KG chain can be approximately mapped onto a DNLS model \cite{DNLS-KG} using $\beta S \approx 6 \lambda \mathcal{H}$. Further analytics will be discussed only in terms of the DNLS chain, since it is then straightforward 
to project to the KG model.

For the DNLS model we measure the spreading of the wave packet by tracking the quantity $n_j \equiv \left|\psi_j\right|^2/S$, hereafter named norm density consistently with the notation of Refs.\cite{flach2009,skokos2009,flach2010,laptyeva2010}. The key quantities that we use to describe the time evolution of the expanding wave packet are the second moment $m_2=\sum_j n_j(X-j)^2$, ($X=\sum_j n_j j$) which quantifies the spatial extent of the wave packet, and the participation number $P=1/\sum_j n_j^2$ which measures the number of significantly populated sites. A combination of these two quantities $\xi=P^2/m_2$, called compactness index, gives a measure of the sparsity of a wave packet \cite{skokos2009}. For the KG we do exactly the same, but replacing the norm density $n_j$ with its counterpart $\mathcal{E}_j/\mathcal{H}$, which is the normalized energy density.

\section{Relevant energy scales}
Neglecting the nonlinear term in Eq.~(\ref{eq:DNLS}) reduces to an eigenvalue problem
\begin{equation}
-A_{\nu,j+1}-A_{\nu,j-1}+\lambda\cos(2\pi\alpha j+\varphi)A_{\nu,j}=E_\nu A_{\nu,j} \, .  \label{eq:AA}
\end{equation}
Here the index $\nu$ labels the different normal modes $A_{\nu,j}$ and eigenvalues $E_\nu$. The coefficient $1/(2\lambda)$ in Eq.~(\ref{eq:KG}) was chosen so that the linear parts of $H_{\rm DNLS}$ and $H_{\rm KG}$ would correspond to the same eigenvalue problem: the linear KG model can then likewise be identically reduced to Eq.~(\ref{eq:AA}), under the substitution $E_\nu = 2 \lambda \left(\omega_\nu^2 - 1/\lambda -1\right)$, where $\omega_\nu$ are the corresponding eigenfrequencies.

Eq.~(\ref{eq:AA}) is also known as the Aubry-Andr\`e model \cite{aubry1980}. The localization properties of this model are well known and extensively studied both analytically and numerically 
\cite{aubry1980, hiramoto1988, jitomirskaya1999, hiramoto1992, modugno2009, larcher2009, larcher2011}.  A transition occurs from an extended regime to a localized regime at $\lambda=2$. For $\lambda<2$ all 
normal modes $A_{\nu,j}$ are extended over the entire lattice, at $\lambda=2$ they are critical, while for $\lambda>2$ they are exponentially localized in the form $A_{\nu,j}\sim e^{-|j-j_\nu|/\ell}$, where $j_\nu$ is 
the localization center and $\ell=1/\ln(\lambda/2)$ is the localization length (notice that it is the same for all the modes) \cite{aubry1980}. Since we are interested in the interplay between localization and nonlinearity, we will focus exclusively on the regime $\lambda>2$.

\begin{figure}[b!]
\begin{center}
\includegraphics[width=0.95\columnwidth]{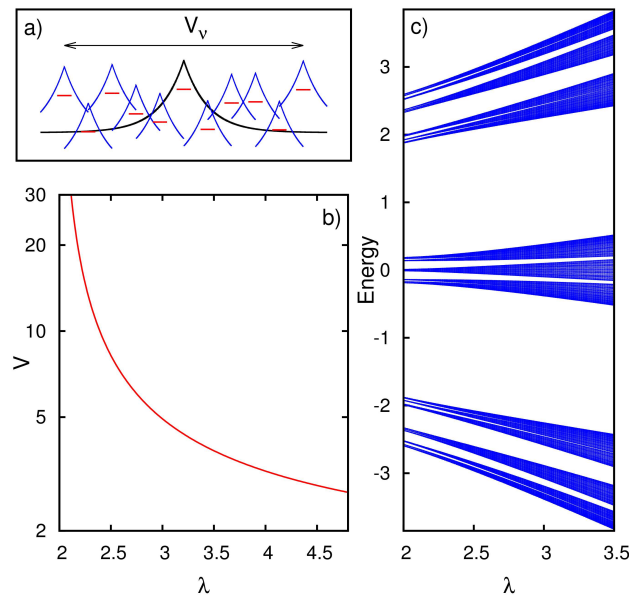}
\end{center}
\caption{(color online) a) Pictorial interpretation of localization volume. A given eigenstate $\nu$ (black line in the center of the box) is assumed to interact only with those eigenstates (blue lines) that lie in a region of size $V_\nu$ around his mean position. The red lines represent the corresponding on-site energies. b) Average localization volume of eigenstates $V$ as a function of the potential strength $\lambda$. 
c) Eigenenergies $E_\nu$ of the linear system obtained from numerical diagonalization of Eq.~(\ref{eq:AA}), as a function of $\lambda$.}
\label{fig:spectrum}
\end{figure}

In order to quantify the spatial extent of a given eigenstate, it is convenient to define a localization volume $V_\nu=1+\sqrt{12 m_{2}^{(\nu)}}$, where $m_{2}^{(\nu)}=\sum_j (X_\nu-j)^2|A_{\nu,j}|^2$ is the second moment of $|A_{\nu,j}|^2$ and $X_\nu=\sum_j j |A_{\nu,j}|^2$ is its center of norm \cite{krimer2010}. The localization volume is used to estimate the number of modes which interact with a given mode $\nu$. We show its meaning schematically in Fig.~\ref{fig:spectrum}$a$. The modes that interact with a given reference mode $\nu$ are those whose center of norm lies in an area $V_\nu$ around it. The average localization volume $V$ is then found by numerically diagonalizing the linear system, calculating $V_\nu$ for each eigenmode, and then averaging over all eigenmodes. A plot of this quantity as a function of the potential strength $\lambda$ is shown in Fig.~\ref{fig:spectrum}$b$.

The spectrum for $\lambda>2$ is purely dense-point, characterized by the presence of an infinite number of gaps and bands.  A plot of the Aubry-Andr\`e model's spectrum as a function of $\lambda$ is shown in 
Fig.~\ref{fig:spectrum}$c$. In this figure, one clearly sees the presence of two major gaps dividing the spectrum in three parts, each of them divided in turn in three smaller parts, and so on \cite{spectrum}. We call these portions of spectrum separated by the largest gaps ``mini-bands''. For our purposes, it is enough to consider a division of the spectrum in $M=3$ or at most in $M=9$ mini-bands. Smaller mini-bands have vanishingly small effects on the time evolution of wave packets.

Let us introduce two energy scales associated with the linear system \cite{flach2009,krimer2010}. The first one, $\Delta$, is the full width of the spectrum, defined as the difference between the largest and the smallest eigenvalues:  $\Delta=\max\{E_{\nu}\}-\min\{E_{\nu}\}$. The second one, $d$, is the mean spacing of eigenvalues within a single mini-band and within the range of a localization volume. Let us explain how we calculate this quantity. We consider a given mini-band and all the eigenstates that lie in it. For each eigenstate $\nu$, we calculate its localization volume $V_\nu$ and then we form the subset of the other eigenstates, $\{\mu\}$, belonging to the same mini-band and interacting with it, namely, those fulfilling the condition $|X_\nu-X_\mu|<V_\nu/2$. The average number of states in the subset can be estimated as $V/M$. Then we calculate the energy spacings within this subset. This procedure is repeated for each eigenstate in the band and the average gives the mean spacing $d$.

\begin{figure}[bt!]
\begin{center}
\includegraphics[width=0.95\columnwidth]{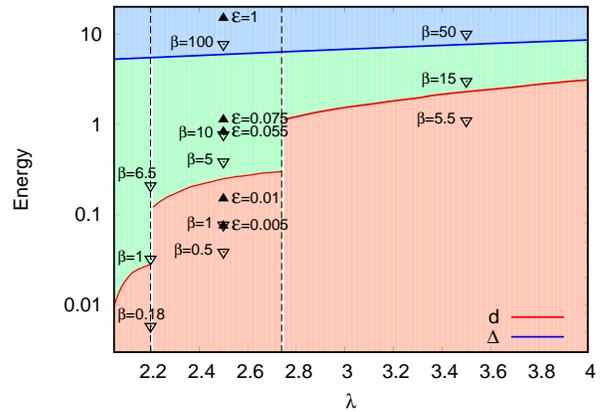}
\end{center}
\caption{(color online) Energy scales $\Delta$ (top blue line) and $d$ (bottom red line) plotted as a function of the potential strength $\lambda$. The empty (downward) and full (upward) triangles correspond to the values of $\delta$ that we have used for the simulations with the DNLS model and with the KG model respectively. Comparing the nonlinear frequency shift $\delta$ with the energy scales $\Delta$ and $d$ one can predict the different spreading regimes of weak chaos ($\delta<d$), strong chaos ($d<\delta<\Delta$) and self-trapping ($\delta>\Delta$). The separation between the three regimes should not be interpreted as a sharp boundary, but as a smooth crossover.}\label{fig:energy_scales}
\end{figure}

The number of mini-bands $M$ to be used in the calculations of $d$ depends on $\lambda$. For a given $\lambda$ we choose $M$ in such a way that the localization volume $V$ satisfies the condition $V/M>2$. This implies that, on average, there are at least two eigenstates within the subset $\{\mu\}$ that we can use to calculate the average energy spacings. We always consider $\lambda>2.1$; therefore it is enough to divide the spectrum at most in nine mini-bands. As $\lambda$ is increased the average localization volume of the eigenstates $V$ decreases -- therefore at some point we have to consider the spectral separation into smaller mini-bands. In practice we consider $M=9$ mini-bands for $2.1\lesssim\lambda\lesssim2.2$, $M=3$ mini-bands for $2.2\lesssim\lambda\lesssim2.75$ and just one band (i.e., the full spectrum) for $\lambda \gtrsim 2.75$.  A plot of the energy scales $\Delta$ and $d$ as a function of $\lambda$ is shown in Fig.~\ref{fig:energy_scales}. The dashed vertical lines represent the values of $\lambda$ where the number of mini-bands changes in the calculation of $d$.

Similarly to the case of disordered systems \cite{flach2009,skokos2009,flach2010}, the scales $\Delta$ and $d$ of the linear spectrum (which are frequencies in the present setting of nonlinear wave equations) must be compared to the frequency shift caused by the nonlinearity. Indeed a single oscillator which satisfies the equation of motion $i\dot{\psi}=V \psi + \beta |\psi|^2\psi$ experiences a nonlinear frequency shift $\delta=\beta |\psi|^2$ away from its linear frequency $V$.  For many oscillators, we can conveniently use the eigenstates of the linear Aubry-Andr\`e model as a decomposition basis of the wave function $\psi_j$: $\psi_j=\sum_\nu \phi_\nu A_{\nu,j}$. Equation (\ref{eq:DNLS}) can then be rewritten for the evolution of the normal mode amplitudes: 
\begin{equation}
i \frac{\partial \phi_\nu}{\partial t}= E_\nu \phi_\nu +\beta \sum_{\nu_1,\nu_2, \nu_3} I_{\nu,\nu_1,\nu_2,\nu_3}\phi_{\nu_1}^*\phi_{\nu_2}\phi_{\nu_3} \label{eq:DNLS_NM}
\end{equation}
where $I_{\nu,\nu_1,\nu_2,\nu_3}$ is an overlap integral involving four normal modes:
\begin{equation}
I_{\nu,\nu_1,\nu_2,\nu_3}=\sum_j A_{\nu,j}A_{\nu_1,j}A_{\nu_2,j}A_{\nu_3,j}.\label{eq:oberlap_integral}
\end{equation}
As discussed in Ref.\cite{laptyeva2010}, one can introduce a norm density also in the normal mode space, $n_\nu = |\phi_\nu|^2$;  as the packet spreads and after averaging over many realization, this quantity becomes almost identical to the norm density $n_j=|\psi_j|^2$ and the frequency shift can be expressed as $\delta \sim \beta n$, where $n$ is a characteristic norm density. In the KG model $\delta$ is proportional to the energy density $\mathcal{E}$ and, within our formalism, can be obtained by the small amplitude mapping.

When $\delta < d$, the mode frequencies in a wave packet are only weakly shifted, and a small fraction of these modes will resonantly and strongly interact with each other. Following the terminology of Refs.~\cite{flach2009,skokos2009,flach2010}, we say that this is a regime of weak chaos. Conversely, when $d < \delta < \Delta$, the mode frequencies in a packet are strongly shifted and almost all of them will resonantly and strongly interact with each other. This is labeled a regime of strong chaos. When finally $\delta > \Delta$, the mode frequencies are shifted so strongly that they are tuned out of resonance with all nonexcited neighbouring modes. An excited mode in this condition may stay localized, i.e., self-trapped, for long or even infinite times. 
The meaning of these regimes will be further clarified in the next section.

\section{Expected spreading regimes}

As one can see in Eq.~(\ref{eq:DNLS_NM}), the presence of nonlinearity in the DNLS model introduces a coupling between eigenstates of the underlying linear spectrum. It has already been observed numerically and experimentally that this leads to a subdiffusive spreading of wave packets, i.e. its second moment grows asymptotically as  $m_2\sim t^\gamma$ with $\gamma<1$ \cite{larcher2009,lucioni2011}. However, a systematic
investigation of the behaviour of the exponent $\gamma$ in different regimes of strong and weak chaos, and self-trapping, have not been done so far. In this section, we approach this issue by first comparing the nonlinear frequency shift $\delta=\beta n$ with the energy scales $\Delta$ and $d$, in such a way as to introduce the different spreading regimes expected to be observed in the subsequent numerical simulations. 

Let us consider an initial wave packet with norm density $n$ and localization volume $L$ larger than the average localization volume of the eigenstates of the linear spectrum, $L\geq V$. If $\delta>\Delta$, nonlinearity is so strong that all the participating normal modes within the wave packet are shifted out of resonance with respect to the non-excited neighbourhood; therefore spreading 
is largely suppressed and a significant part of the wave packet remains self-trapped \cite{self-trapping}. If instead $\delta<\Delta$, we are no longer in the self-trapping regime and can distinguish two sub-cases: on one hand, when $\delta>d$, strong chaos is realized; that is, all the modes in the packet are resonantly interacting with each other, thus producing an efficient spreading. On the other hand, when $\delta<d$, weak chaos is obtained: only a fraction of modes interact resonantly -- the localization is still destroyed, but spreading is slower. 

If $L<V$ the estimate of the self-trapping transition is done as before, that is by comparing $\delta=\beta n$ with the spectrum width $\Delta$. If self-trapping is avoided, however, the wave packet initially spreads also in absence of nonlinearity, eventually filling the localization volume $V$. Consequently the initial norm density $n$ is reduced to $\tilde{n}\approx nL/V$, due to linear time evolution -- the relevant 
nonlinear frequency shift must now be calculated by using this reduced density $\tilde{n}$. Apart from this detail, which originates from the initial dynamics at short times, the asymptotic spreading regimes are the same as before. 

Studies performed on random systems have shown that the basic mechanism that destroys localization is the presence of resonances in mode-mode interaction \cite{flach2009,skokos2009,flach2010}. This leads to chaotic dynamics within a part of the wave packet, and to a subsequent subdiffusive spreading. Here we apply the same theory to the case of quasiperiodic systems. For the reader interested in the formal details of the theory, we refer to the previous articles \cite{skokos2009,flach2010}.

Let us estimate the number of resonant modes in the packet, which is a key quantity that determines the type of spreading behaviour. According to Eq.~(\ref{eq:DNLS_NM}), due to nonlinearity, the evolution of a given normal mode is affected by any three (triplet) modes. The coupling is the largest if the triplet modes have large amplitudes and if the overlap integrals are large, i.e., if the triplet modes are close enough in space to the given normal mode. Some of these triplet modes may affect the dynamics of the chosen mode $\nu$ strongly, some weakly. To distinguish these triplet groups, we follow \cite{flach2010} and apply perturbation theory to first order in $\beta$. It follows that the amplitude of a normal mode $\nu$ inside the wave packet is changed by a given triplet of other wave packet modes $\vec{\mu}=\{\mu_1,\mu_2,\mu_3\}$ as
\begin{equation}
|\phi_\nu^{(1)}|=\beta\frac{\sqrt{n_{\mu_1} n_{\mu_2} n_{\mu_3}}}{R_{\nu,\vec{\mu}}} \label{eq:perturbation}
\end{equation}
where
\begin{equation}
R_{\nu,\vec{\mu}}\sim \left|\frac{E_\nu+E_{\mu_1}-E_{\mu_2}-E_{\mu_3}}{I_{\nu,\mu_1,\mu_2,\mu_3}}\right| \, .
\label{eq:R}
\end{equation}
From now on, we assume that all the modes that belong to the packet (i.e., that are located between the two exponential tails of the wave packet) have the same norm density equal to $n$. The perturbation approach breaks down and resonance sets in when $\sqrt{n}<|\phi_\nu^{(1)}|$. Substituting 
Eq.~(\ref{eq:perturbation}) for $\phi_\nu^{(1)}$ one can rewrite the last inequality as
\begin{equation}
R_{\nu,\vec{\mu}} < \beta n.	
\label{eq:condition}
\end{equation}
This expression tells us that the resonance condition, for a given normal mode $\nu$, is fulfilled if there is at least one triplet of modes $\vec{\mu}$ that satisfies inequality (\ref{eq:condition}).

\begin{figure}[t!]
\begin{center}
\includegraphics[width=0.95\columnwidth]{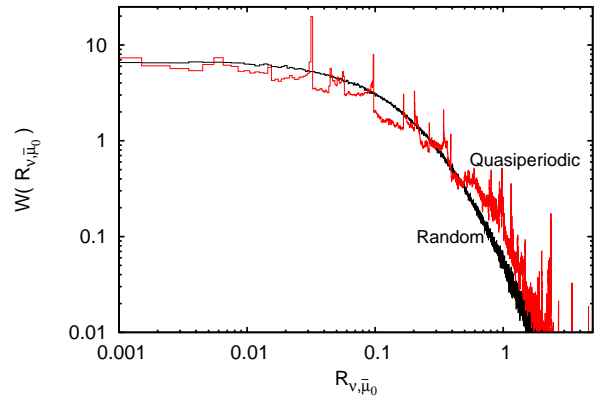}
\end{center}
\caption{(color online) Comparison between the probability density function $\mathcal{W}(R_{\nu,\vec{\mu_0}})$ of the quasiperiodic DNLS model and of the random DNLS model. For the quasiperiodic case, $\lambda=2.5$, while for the random case, we choose a disorder strength that gives a similar localization length.} \label{fig:resonance}
\end{figure}

The probability for the onset of a resonance can therefore be calculated with the following statistical numerical analysis \cite{flach2009,krimer2010}. For a given normal mode $\nu$, we define $R_{\nu,\vec{\mu_0}}= \min_{\vec{\mu}} \{R_{\nu,\vec{\mu}}\}$. Collecting $R_{\nu,\vec{\mu_0}}$ for many modes and many values of the phase $\varphi$, we find the probability density distribution $\mathcal{W͑}(R_{\nu,\vec{\mu_0}})$.  From this quantity we can calculate the probability $\mathcal{P}$ for a mode, with norm density $n$, to be resonant with at least one triplet of other modes at a given value of the interaction parameter $\beta$. This is obtained by integrating $\mathcal{W}(R_{\nu,\vec{\mu_0}})$ from zero to $\beta n$
\begin{equation}
\mathcal{P}=\int_0^{\beta n} \mathcal{W}(R)\, dR.\label{eq:resonance_probability}
\end{equation}
An example of probability density $\mathcal{W}(R_{\nu,\vec{\mu_0}})$ for $\lambda=2.5$ is shown in Fig.~\ref{fig:resonance} (red line). For comparison we also show the same quantity for the random DNLS model (black line), as discussed in \cite{flach2009,krimer2010}. Except for fine structures, like small sharp peaks appearing in the quasiperiodic case, the overall behaviour is qualitatively very similar in the two cases. In particular, the probability density $\mathcal{W}$ tends to a finite constant value $C$ when $R_{\nu,\vec{\mu_0}}\rightarrow 0$. As a consequence, for small values of $\beta n$, a non-zero fraction of modes in the packet is resonant. The probability to be resonant is given by $\mathcal{P}\sim C\beta n$, thus we are in the weak chaos regime. For large values of $\beta n$, instead all the modes interact resonantly and $\mathcal{P}=1$; we are then in the strong chaos regime. 

\begin{figure*}[t!]
\begin{center}		
\includegraphics[width=1.58\columnwidth,keepaspectratio,clip]{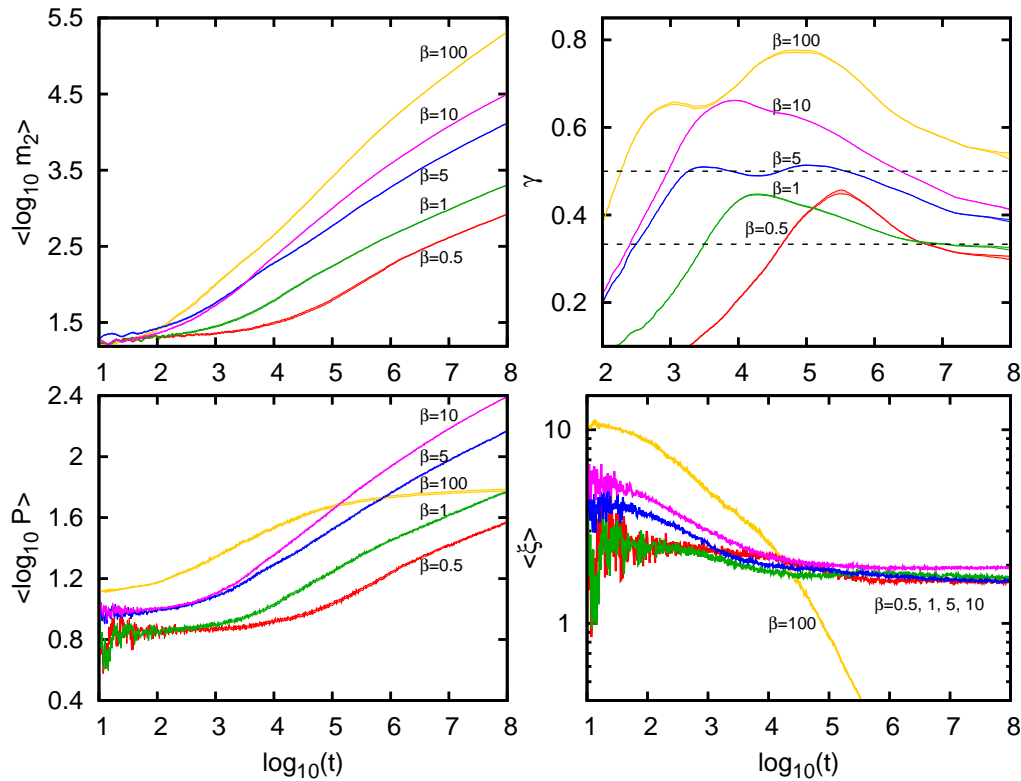}
\end{center}
\caption{(color online) Numerical results obtained by integrating the DNLS equations of motion (\ref{eq:DNLS}). The time evolution of $\left\langle \log_{10} m_2\right\rangle$ (left panel, top), $\gamma$ (right panel, top), $\left\langle \log_{10} P \right\rangle$ (left panel, bottom), and $\left\langle \xi\right\rangle$ (right panel, bottom) is shown versus $\log_{10}t$ for different values of the nonlinear parameter $\beta=0.5, 1, 5, 10, 100$. The initial wave packet in all simulations is a square distribution with $L=13$ and the potential strength is $\lambda=2.5$. In the top right panel the two dashed lines correspond to theoretically predicted power laws $\gamma=1/3$ and $\gamma=1/2$. The width of the lines for the quantities $\langle\log_{10}m_2\rangle$, $\langle\log_{10} P\rangle$ and $\gamma$ represents the statistical error, which depends on time and on the number of realizations. In most cases the statistical error is smaller than the resolution of the figure. All quantities are dimensionless.}
\label{fig:time_evolution}
\end{figure*}

Following the reasoning presented in \cite{flach2010}, this implies that also in the quasiperiodic case, as in disordered systems, we may expect to find $m_2\sim t^{1/3}$ in the weak chaos regime and $m_2\sim t^{1/2}$ in the strong chaos regime. Note the strong chaos regime can only exist as a transient regime: as the wave packet spreads, its norm density $n$ decreases, and eventually will reach a situation where $\beta n<d$. At this point, a crossover from strong to weak chaos is expected to occur during the time evolution \cite{laptyeva2010}. 

Let us finally stress that the ``transition lines'' that we have introduced by comparing the nonlinear frequency shift with the typical energy scales of the linear spectrum do not define sharp phase transitions 
between different spreading regimes. Instead, we may expect to see a relatively smooth crossover, such that the regimes of self-trapping, strong chaos and weak chaos should be clearly identified only far from the transition lines.

\section{Time evolution} 

We perform extensive numerical simulations solving Eqs.~(\ref{eq:DNLS}) and (\ref{eq:KG}) for different sets of parameters $\left\{\lambda,\beta\right\}$ and $\left\{\lambda,\mathcal{E}\right\}$, respectively. For each choice of parameters we average over $N$ different realizations of the quasiperiodic potential obtained by randomly changing the phase shift $\varphi$. For initial conditions, we use compact wave packets that lie in the center of our computational box, taking care that during the time evolution the wave packet never reaches the box boundaries. The number of realizations considered varies between $100$ and $500$ and the number of lattice sites between $200$ and $2000$. To solve the equations of motion, we use symplectic integration schemes of the SABA family \cite{laskar2001,skokos2009}
that allow us to reach large integration times with good accuracy \cite{precision}.

In order to quantify the type of subdiffusive behaviour, we calculate the exponent $\gamma$ by considering the logarithm of the second moment $\log_{10} m_2$ for different realizations of the potential. We compute the average value $\langle \log_{10} m_2\rangle$ and its statistical error, given by the standard deviation divided by the square root of the number of realizations $N$. Then the value of $\gamma$ at a given time $t$ is calculated by applying a linear fitting procedure to the curve $\langle \log_{10} m_2\rangle$ within a fixed time interval around $\log_{10} t$. By repeating this procedure at different $t$, we extract the behaviour of $\gamma$ as a function of time and its relative statistical error.

\subsection{Results of the DNLS model}

Let us first show our results for the DNLS model. For the initial wave packet, we choose a square shaped distribution which equally populates $L$ lattice sites with norm density $n_j=n=1/L$. In Fig.~\ref{fig:time_evolution} we present a representative set of simulations for $\lambda=2.5$. We choose $L=13$, which gives an initial localization volume larger than $V$. The different panels show the time evolution of the second moment $\langle\log_{10}m_2\rangle$, the spreading exponent $\gamma$, the participation ratio $\langle\log_{10} P\rangle$, and the compactness index $\langle \xi \rangle$. The width of the curves for
$\langle\log_{10}m_2\rangle$, $\langle\log_{10} P\rangle$ and $\gamma$ corresponds to the statistical error. The values of the nonlinear frequency shift $\delta$ induced by the initial wave packets used in these 
simulations are shown in Fig.~\ref{fig:energy_scales} (empty downward triangles) in order to compare them to the relevant energy scales $\Delta$ and $d$. 

In all simulations we observe that nonlinearity causes the wave packet to spread. The spreading starts earlier when $\beta$ is larger. We find that the spreading is always subdiffusive ($\gamma<1$), confirming 
the result of previous works \cite{larcher2009, lucioni2011}. Subdiffusion is seen both in the second moment $m_2$ and in the participation ratio $P$, except for the largest value of $\beta$ (yellow curves 
in Fig.~\ref{fig:time_evolution}). In the latter case, $P$ saturates to a constant value after a transient time -- a clear signature of self-trapping. This observation of self-trapping only for $\beta=100$ is consistent with the energy scale arguments schematically represented in Fig.~\ref{fig:energy_scales}. In the absence of self-trapping, the compactness index $\xi$ saturates to a constant value, indicating that the wave packet spreads but does not become more sparse. Conversely, in the presence of self-trapping the central part of the wave packet remains spatially trapped while its tails keep expanding, thus resulting in a wave packet that becomes more sparse during the evolution -- nicely quantified by the compactness index which decreases to zero. We notice that the portion of packet that is expanding is characterized by a value of $\gamma$ larger than $1/2$. After an initial increase, $\gamma$ reaches a maximum and then decreases to smaller values. In this regime, the evolution is rather complex. A similar behaviour was previously obtained also in random systems \cite{laptyeva2010,bodyfelt2011}. The transient large values of $\gamma$ may be due to a nontrivial interacting mechanism that takes place between the expanding part and the self-trapped portion, resulting in faster spreading - an effect labeled as ``overshooting''.

For the lowest values of $\beta$ the energy scale arguments suggest the occurrence of weak chaos. Indeed for $\beta=0.5$ and $1$ the exponent saturates asymptotically around the theoretical value $\gamma=1/3$ (red and green curves in Fig.~\ref{fig:time_evolution}), as expected. It is worth mentioning that this asymptotic exponent is the same as in random systems \cite{flach2010,laptyeva2010}; meaning that the mechanism leading to destruction of exponential localization is rather universal. 

In difference to the random case, here during the time evolution, the value of $\gamma$ temporarily increases above $1/3$, eventually reaching its asymptote only at longer times. This is an overshooting similar to the one that we have discussed above for the self-trapping regime, but occurring also for weaker nonlinearity. This effect is unique to the quasiperiodic system and is likely due to the presence of an infinite number of mini-bands and gaps in the linear spectrum of the Hamiltonian, which causes a temporary self-trapping of portions of the expanding wave packet in one or more energy gaps between mini-bands. This partial self-trapping is different from the self-trapping that occurs when $\delta>\Delta$, where all the packet modes are simultaneously shifted out of resonance.  For this reason partial self-trapping is not detectable as a saturation of the participation number $P$ and can only be seen indirectly as an overshooting in the exponent $\gamma$.  

The two simulations for  $\beta=5$ and $10$ lie in a range of energy were we expect to see strong chaos  (blue and magenta curves in Fig.~\ref{fig:time_evolution}). As already said in the previous section, the
strong chaos regime is transient: one should find  a value of $\gamma$ around $1/2$ for a few decades of time, eventually decreasing towards the asymptotic value $1/3$.  The two corresponding curves in 
Fig.~\ref{fig:time_evolution} indeed exhibit a behaviour which qualitatively agrees with this expectation. The value of $\gamma$ first rises up to $1/2$, oscillates around this value and then starts to decrease as predicted. However, especially for large $\beta$, we also observe values of $\gamma$  larger than $1/2$. As in the weak chaos regime, this overshooting again is evidence of partial self-trapping. Its mechanism is also transient and occurs in the same time intervals where strong chaos is expected. For this reason, while weak chaos is clearly observed in our simulations, strong chaos and partial self-trapping tend to overlap, thus producing a more complex evolution of the wave packet in quasiperiodic systems than in random systems. 

\begin{figure}[t!]
\begin{center}
\includegraphics[width=\columnwidth]{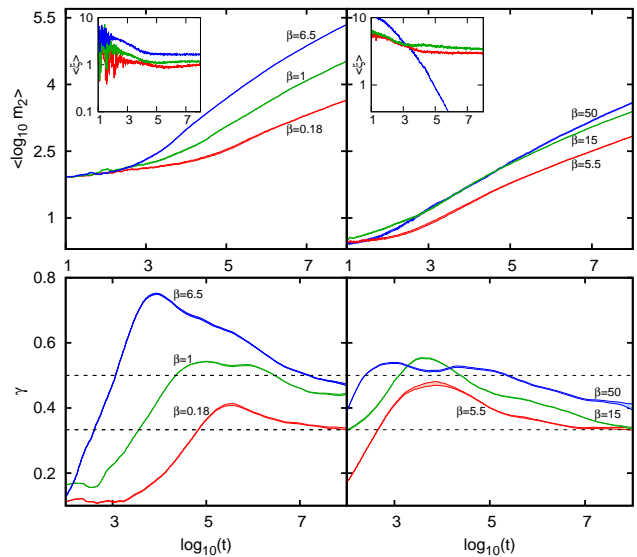}
\end{center}
\caption{(color online) Average logarithm of the second moment of the expanding wave packet, $\langle \log_{10} m_2 \rangle$ and spreading exponent, $\gamma$ for $\lambda=2.2$ (left plots) and $\lambda=3.5$ (right plots).For $\lambda=2.2$, the initial wave packet has width $L=31$ and we consider $\beta=0.18$ (lower red curves),$1$ (mid green curves) and $6.5$ (upper blue curves). For $\lambda=3.5$, 
the initial wave packet has width $L=5$ and we consider $\beta=5.5$ (lower red curves), $15$ (mid green curves), and $50$ (upper blue curves).
The width of the lines represents the statistical error as in Fig.~\ref{fig:time_evolution}. Insets: average compactness index of the expanding wave packet $\langle \xi \rangle$ for the same sets of simulations.}
\label{fig:time_evolution_many_lambda}
\end{figure}

\begin{figure*}[t!]
\includegraphics[width=1.58\columnwidth,keepaspectratio,clip]{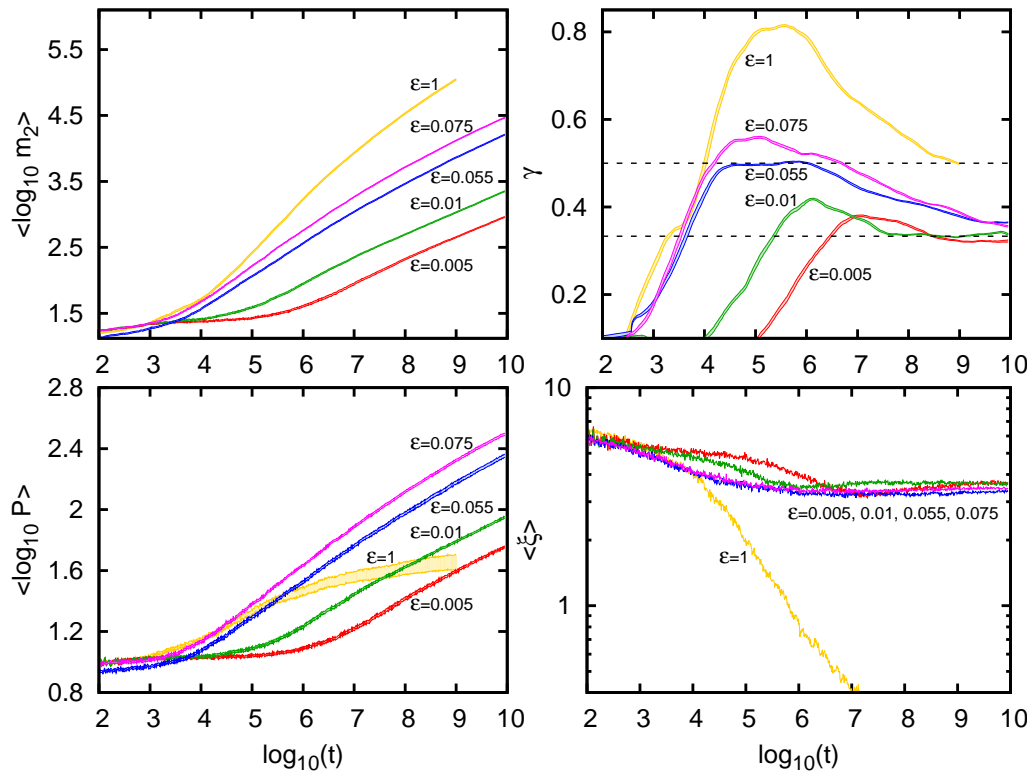}
\caption{(color online) Numerical results obtained by integrating the KG equations of motion (\ref{eq:KG}). The time evolution of $\left\langle \log_{10} m_2\right\rangle$ (left panel, top), $\gamma$
(right panel, top), $\left\langle \log_{10} P \right\rangle$ (left panel, bottom), and $\left\langle \xi \right\rangle$ (right panel, bottom) is shown versus $\log_{10}t$. The parameters are $\left\{\lambda, \mathcal{E}\right\}= \left\{2.5, 0.005\right\}, \left\{2.5, 0.01\right\}, \left\{2.5, 0.055\right\}, \left\{2.5, 0.075\right\}, \left\{2.5, 1.0\right\}$. We used an initial wave packet with width $L = 13$ for $\mathcal{E}=0.005,\,0.01,\,0.075,\,1$ and $L=11$ for $\mathcal{E}=0.075$. The width of the lines for the quantities $\langle\log_{10}m_2\rangle$, $\langle\log_{10} P\rangle$ and $\gamma$ represents the statistical error as in Fig.~\ref{fig:time_evolution}. In the top right panel the two dashed lines correspond to theoretically predicted power laws $\gamma=1/3$ and $\gamma=1/2$.}
\label{fig:num_KG}
\end{figure*}

In Fig.~\ref{fig:time_evolution_many_lambda} we show the results of simulations for $\lambda=2.2$ and $\lambda=3.5$; the corresponding values of nonlinear frequency shift are reported as triangles in Fig.~\ref{fig:energy_scales}. The values of $L$ are $L=31$ for $\lambda=2.2$ and $L=5$ for $\lambda=3.5$, both larger than $V$. For $\{\lambda,\beta\}=\{2.2,0.18\}$ and $\{\lambda,\beta\}=\{3.5,5.5\}$ energy scale arguments predict weak chaos. We indeed find a spreading exponent which approaches asymptotically the value $1/3$. For $\{\lambda,\beta\}=\{2.2,1\}$, $\{\lambda,\beta\}=\{2.2,6.5\}$ and $\{\lambda,\beta\}=\{3.5,15\}$ the predicted behaviour is either strong chaos or a regime in between strong and weak chaos.  What we observe numerically is a growth of the spreading exponent $\gamma$ up to $1/2$ and even to larger values, followed by a decrease towards $1/3$. In most cases, our simulations show a significant overshooting due to partial self-trapping. It is worth mentioning that this effect is larger for weaker disorder strength $\lambda$, consistent with the fact the linear spectrum exhibits larger mini-gaps in this regime (see Fig.~\ref{fig:spectrum}). Finally for $\{\lambda,\beta\}=\{3.5,50\}$, we observe self-trapping, as expected. 

In conclusion, from the analysis of the results of the DNLS model for different values of $\lambda$ we find that the energy scale arguments and the model discussed in Section IV correctly explain the overall trend of the numerical simulations and the separation between different spreading regimes in the parameter space.

\subsection{Results of the KG model}
Due to the existence of a mapping between KG and DNLS, we expect to observe the same spreading regimes in the two models. This has been already proven in purely random systems where the two models reveal similar qualitative results in a wide range of parameters \cite{flach2009,skokos2009,flach2010,laptyeva2010,bodyfelt2011}. Despite this similarity, the study of the KG model remains interesting for at least two reasons. On one hand, it allows for testing the generality of the result in a case where there is just one conserved quantity. This is highly nontrivial -- especially for self-trapping -- for which rigorous results have been recently derived only in the case of Hamiltonians conserving both energy and norm \cite{kopidakis2008}. On the other hand, the KG model is advantageous from a numerical point of view. The fact that there is just one conserved quantity results in two orders of magnitude faster integration speed within the same integration error.

Similarly to what was done for the DNLS model, we initially set the compact wave packets to span a width $L=13$ (unless otherwise stated) centered in the lattice, such that each site has equal energy $\mathcal{E}_j = \mathcal{E}=\mathcal{H}/L$. This is implemented by setting initial momenta of $p=\pm \sqrt{2 \mathcal{E}}$ with randomly assigned signs and zero coordinates. The values of initial energy densities $\mathcal{E}$ are chosen to give expected spreading regimes of asymptotic weak chaos, intermediate strong chaos, and dynamical crossover from strong chaos to the slower weak chaos subdiffusive 
spreading \cite{flach2010}. 

The results of the time simulations are shown in Fig.~\ref{fig:num_KG}, while the expected spreading regimes are given in Fig.~\ref{fig:energy_scales} (full upward triangles) \cite{freq_shift}. As one can see by comparing Fig.~\ref{fig:num_KG} with Fig.~\ref{fig:time_evolution}, the qualitative behaviour of the two models is rather similar. After initial transients, which increase with decreasing nonlinearity, all KG simulations reveal subdiffusive growth of the second moment $m_2$ according to power law $m_2 \sim t^\gamma$ with $\gamma<1$. If self-trapping is avoided, all simulations show a similar subdiffusive behaviour for the participation numbers; moreover, the wave packets remain compact as they spread, since compactness indices at the largest computational times saturate around a constant $\left\langle \xi \right\rangle \approx 3.5 \pm 0.25$. For the two smallest values of initial energy density $\mathcal{E}=0.05$ and $\mathcal{E}=0.01$, the characteristics of the weak chaos regime are observed, namely, the exponent $\gamma$ 
saturates around  $1/3$ (red and green curves in Fig.~\ref{fig:num_KG}) after a transient time. We stress that the only difference from the purely random systems is the overshooting phenomenon at transient times. This effect is an inherent property of quasiperiodic systems which inevitably manifests itself in all spreading regimes, while in the disordered case it was shown to occur only in the regime of self-trapping 
\cite{laptyeva2010,bodyfelt2011}.

For the two energy densities $\mathcal{E} = 0.055$ and $0.075$ we suggest strong chaos, with characteristics similar to the DNLS case. The simulation with $\mathcal{E} = 0.055$ (blue curves in Fig.~\ref{fig:num_KG}) 
indeed exhibits the typical behaviour of the strong chaos scenario: the characteristic exponent $\gamma$ increases up to predicted value $1/2$ and remains so for about two time decades, followed by a crossover with $\gamma$ decreasing to the weak chaos dynamics. There is also another possibility for larger $\mathcal{E}=0.075$, when intermediate strong chaos is masked due to partial self-trapping (magenta curves in Fig.~\ref{fig:num_KG}). Thus, $\gamma$ shows values larger then $1/2$ but still with subsequent decay to slower subdiffusion. Here, we would like to strongly emphasize that none of the simulations exhibit pronounced deviations from strong or weak chaos regimes of spreading, i.e. long-lasting overshooting with $\gamma>1/2$, or significant slowing down to values $\gamma < 1/3$.

Finally, for $\mathcal{E}=1.0$ the dynamics enter the self-trapping regime, as our theory predicts. There a major part of the initial excitation stays localized, while the remainder spreads (yellow curves in Fig.~\ref{fig:num_KG}). The participation numbers, therefore, do not grow significantly and $\left\langle\log_{10}P\right\rangle$ starts to level off at large time (Fig.~\ref{fig:num_KG}, left panel, bottom, yellow curve). In contrast, the small spreading portion yields a continuous increase of the second moment $m_2$  (Fig.~\ref{fig:num_KG}, left panel, top, yellow curve), which initially is characterized by large values of $\gamma > 1/2$ (howbeit, for larger time $\gamma$ decreases). Consequently, the compactness index $\left\langle\xi\right\rangle$ (Fig.~\ref{fig:num_KG}, right panel, bottom, yellow curve) drops down to small values indicating deep self-trapping regime. Note that a similar behaviour has been observed before in purely random systems \cite{laptyeva2010,bodyfelt2011}. Unusually large 
values of $m_2$ can be explained by local trapping-detrapping processes in the evolving wave packet. The corresponding dynamics is in strong non-equilibrium -- its theoretical description has yet to
be developed.

\subsection{Role of the shape of the initial wave packet}

In this subsection we show that the results discussed so far do not depend on the shape of the initial wave packet. Besides its theoretical interest, this issue is also relevant from the point of view of experiments, where it is not always possible to design the wave packets at will.

In the previous sections, we have used a square distribution as the initial wavepacket. Now, inspired by the experiments with ultracold atoms, we consider initial wave packets with the shape of a Gaussian distribution or a Thomas-Fermi (TF) distribution. The Gaussian wave packet centered around the site $j=0$ has the form  
\begin{equation}
\psi_j(0)=C_1 e^{-\frac{j^2}{2\sigma^2}}, 
\end{equation}
where $\sigma$ is a parameter controlling the width of the packet while $C_1$ is a constant factor that can be determined by using the normalization condition $\sum_j |\psi_j|^2=1$. A Thomas-Fermi wave packet is instead defined by 
\begin{equation}
\psi_j(0)=C_2 \sqrt{1-\frac{j^2}{R^2}} 
\end{equation}
in the region where $|j|<R$ and $\psi_j=0$ otherwise. The parameter $R$ is the Thomas-Fermi radius characterizing the width of the distribution, while the constant $C_2$ is a normalization factor. 
These two distributions are of interest when considering ultracold bosons initially released from an harmonic trap in the Gross-Pitaevskii regime \cite{dalfovo1999}. 

\begin{figure}[t!]
\begin{center}		
\includegraphics[width=1\columnwidth]{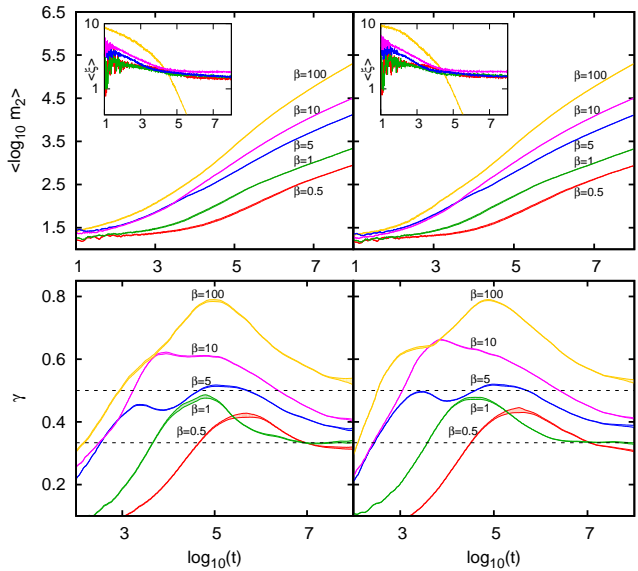}
\end{center}
\caption{(color online) Average logarithm of the second moment of the expanding wave packet, $\langle \log_{10} m_2 \rangle$ and spreading exponent, $\gamma$ as a function of time for different nonlinearities $\beta=0.5,1,5,10,100$. The disorder strength is $\lambda=2.5$ in all simulations. As an initial condition, we have used a Gaussian wave packet with $\sigma=5$ (left plots) and a TF distribution 
with $R=7.50$ (right plots). The width of the lines represents the statistical error as in Fig.~\ref{fig:time_evolution}. Insets: average compactness index of the expanding wave packet $\langle \xi \rangle$ for the same sets of simulations.} \label{fig:different_packets}
\end{figure}

In Fig.~\ref{fig:different_packets} we show the time evolution in the DNLS model of the second moment of the expanding wave packet, $\langle \log_{10} m_2 \rangle$ (top row) and of the spreading exponent, $\gamma$ (bottom row), using initially a Gaussian (left column) and a TF (right column) wave packet distribution. In the insets we also show the compactness index $\langle \xi \rangle$, in order to identify the self-trapping regime. We choose the width of the initial distributions ($\sigma$ and $R$) so that the nonlinear frequency shift is similar to the one already used for the simulations in Fig.~\ref{fig:time_evolution} \cite{n_average}. In particular we use $\sigma=5$ and $R=7.5$, yielding a nonlinear frequency shift $\delta\approx\beta/13$. The values of $\beta$ used in Fig.~\ref{fig:different_packets} are the same as those previously considered. 

From the comparison between the results of Fig.~\ref{fig:different_packets} and Fig.~\ref{fig:time_evolution}, we can conclude that the shape of the initial wave packet does not affect the overall behaviour of the time evolution, nor its interpretation in terms of regimes of weak and strong chaos, self-trapping, and overshooting. This suggests the results that we have obtained are rather general and that the 
nonlinear frequency shift $\delta$ is the only key parameter controlling the dynamics of the wave packet.  

\subsection{Application to cold atoms}

The DNLS model can be used to simulate the dynamics of bosons in optical lattices at zero temperature \cite{trombettoni2001} and in the tight-binding regime, where the DNLS equation corresponds to a discretized version of the Gross-Pitaevskii (GP) equation for the dynamics of a Bose-Einstein condensate in the single-band approximation. The validity of this mean-field theory is not ensured for those dynamical regimes where GP predicts chaos \cite{validityGP}, which can be viewed as a signature of a large depletion of the condensate. For this reason, in the presence of disorder the theory fails to predict the long time evolution of observables directly related to small scale fluctuations and long-range coherence. However, for coarse-grained observables, like the width of the wave packet in real and momentum space, or the participation number, the predictions of the theory remain very good even in regimes where the depletion is expected to be large, long after the random fluctuations prevent the prediction of fine scale structures. This has been recently shown in Ref.~\cite{brezinova12} by comparing the predictions of the GP equation with one beyond mean-field theory in numerical simulations within timescales of the order of typical experiments with cold atoms and long enough to observe the effects of depletion and chaotic dynamics. Indeed our analysis is essentially based on coarse-grained observables. In addition, for each set of parameters we also average over many realizations and this extends the validity of the present approach even for longer times, as any residual dependence on small scale fluctuations is further suppressed by the averaging procedure.

When applied to bosons expanding in bichromatic optical lattices, our results provide a consistent interpretation of the experimental data of Ref.~\cite{lucioni2011}, where a Bose-Einstein condensate, initially confined in an harmonic trap, is let free to expand in a bichromatic potential. In our dimensionless units, the expansion lasts for times of the order of $10^4$ (see \cite{larcher2009} for more details) and the width of the atomic cloud increases up to $50-100$ lattice sites. In this experiment a subdiffusive spreading is observed with exponents $\gamma$ significantly larger than $1/3$ already for weak nonlinearities and even larger than $1/2$ for larger nonlinearities \cite{exponents}. Our work suggests that such large values of $\gamma$ can be explained in terms of a transient overshooting caused by partial self-trapping in mini-bands.

\section{Summary and Conclusions}

In this work we have considered the problem of the interplay between localization and interaction in one-dimensional quasiperiodic systems. We have investigated the expansion of initially localized wave packets in two 
different quasiperiodic models, DNLS and KG. We have confirmed that interaction destroys localization, giving rise to a subdiffusive growth of the second moment of the wave packet ($m_2\sim t^\gamma$ with $\gamma<1$). 

We have interpreted the spreading process in terms of resonances in the mode-mode coupling.  In particular, we have identified the different spreading regimes of self-trapping, strong chaos and weak chaos by comparing the frequency shift induced by the nonlinearity with the energy scales extracted from the spectrum of the underlying linear system. For weak and strong chaos regimes we have also predicted the expected spreading exponents $\gamma=1/3$ and $\gamma=1/2$ respectively.  

We have performed numerical simulations, which last for much longer times than existing simulations, and we have averaged our results over many realizations. This gave us the possibility to accurately calculate the 
spreading exponent $\gamma$ and observe the weak chaos regime. A key difference with respect to random systems \cite{flach2009,flach2010} is the occurrence of transient overshooting regimes that we interpret as due to the peculiar structure of the linear spectrum of the quasiperiodic system, which is separated into mini-bands. These mini-bands are responsible for mechanisms of partial self-trapping. Signatures of strong chaos have also been observed, but  the temporal overlap of strong chaos and partial self-trapping makes the analysis of the spreading more complex than for random systems. We have also verified that our main results do not depend on the details of the shape of the initial wave packet. This suggests that the nonlinear frequency shift, $\delta$, is the only parameter that controls the dynamics.

Finally, our results provide a consistent interpretation of the subdiffusive spreading observed in experiments with ultracold atoms propagating in bichromatic optical lattices \cite{lucioni2011}.

\begin{acknowledgments}
M.L. thanks the Max Planck Institute for the Physics of Complex Systems, Dresden, for their hospitality. We are indebted to G. Modugno and E. Lucioni for fruitful discussions. This work has been supported by ERC through the QGBE grant and by the UPV/EHU under program UFI 11/55.
\end{acknowledgments}


\begin{thebibliography}{99}
\bibitem{alpha} We note that our choice of $\alpha$ is equivalent to the more commonly used value of the inverse of the golden mean $(\sqrt{5}-1)/{2}$.
\bibitem{spectrum} An intuitive understanding of this band structure can be given, following an heuristic argument. The wavelength of the potential $V_j$ is $1/|\alpha|=2.618\dots$ An
effective wavelength equal to an integer number $q$ would correspond to a separation in exactly $q$ bands. Our value of $1/|\alpha|$ lies between two and three, so that the band structure has neither
two nor three bands, but three main bands with an internal structure of sub-bands.
\bibitem{precision} The numerical accuracy of our calculation is controlled by checking the conservation of the energy $\mathcal{H}$ and the norm $S$ of the expanding wave packet. The error is always
kept smaller than $10^{-2.5}$. For the integration we used time steps between $0.1$ and $0.05$.
\bibitem{freq_shift} In order to compare the nonlinear frequency shift $\delta\sim\mathcal{E}$ of the KG model with the energy scales of the DNLS model $\Delta$ and $d$, we use the approximate
mapping $\beta n \approx 6\lambda\mathcal{E}$ at $t=0$. Therefore, the quantity that is plotted in Fig.~\ref{fig:energy_scales} for the KG model is $6\lambda\mathcal{E}$.
\bibitem{n_average} For Gaussian and TF distributions we identify the nonlinear frequency shift with the quantity $\beta \sum_j |\psi_j|^4$, which is also identical (up to a prefactor) with the mean-field interaction energy in Eq.~(1).
\end{thebibliography}
\end{document}